\documentclass[12pt]{article}
\usepackage{graphicx}
\usepackage{amssymb}
\usepackage{color} 
\usepackage{wrapfig}
\usepackage{amsmath}
\usepackage{booktabs}
\renewcommand{\thefootnote}{*\arabic{footnote}}

\makeatletter
\@addtoreset{equation}{section}
\makeatother

\newcommand{\abs}[1]{\left| #1 \right|}

\begin{document}
\begin{titlepage}

\begin{flushright}
IUHET-534 \\
Cavendish-HEP-09/18 \\
DAMTP-2009-66 \\
\end{flushright}

\vspace{1ex}

\begin{center}

{\large \bf Spontaneous CP violation in E$_{\text 6}$ SUSY GUT\\
with SU(2) flavor and anomalous U(1) symmetries}

\vspace{3ex}

\renewcommand{\thefootnote}{\alph{footnote}}

M. Ishiduki$^1$, 
S.-G. Kim$^{23}$\footnote{kimsg@indiana.edu},
N. Maekawa$^1$\footnote{maekawa@eken.phys.nagoya-u.ac.jp}, and
K. Sakurai$^4$\footnote{sakurai@hep.phy.cam.ac.uk}

\vspace{4ex}
$^1${\it Department of Physics, Nagoya University,\\
Nagoya 464-8602, Japan}\\

\vspace{2ex}
$^2${\it Department of Physics, Tohoku University,\\
Sendai 980-8578, Japan}\\

\vspace{2ex}
$^3${\it Department of Physics, Indiana University,\\
Bloomington IN 47405, USA}\\


\vspace{2ex}
$^4${\it DAMTP, Wilberforce Road, Cambridge, CB3 0WA, UK\\
Cavendish Laboratory, JJ Thomson Avenue, Cambridge, CB3 0HE, UK }\\

\end{center}

\renewcommand{\thefootnote}{\arabic{footnote}}
\setcounter{footnote}{0}
\vspace{6ex}


\begin{abstract}

We construct a model of spontaneous CP violation in E$_{\text 6}$ supersymmetric grand unified theory.
In the model, we employ an SU(2)$_\text{F}$ flavor symmetry and an anomalous U(1)$_{\text A}$ symmetry.
The SU(2)$_\text{F}$ flavor symmetry is introduced to provide the origin of hierarchical structures of Yukawa coupling and to ensure the universality of sfermion soft masses.
The anomalous U(1)$_{\text A}$ symmetry is introduced to realize the doublet-triplet mass splitting, to provide the origin of hierarchical structures of Yukawa couplings, and to solve the $\mu$ problem.
In the model, CP is spontaneously broken by the SU(2)$_\text{F}$  breaking in order to provide a Kobayashi-Maskawa phase and to evade the supersymmetric CP problem.
However, a naive construction of the model generally leads to unwanted outcome, Arg$[\mu b^*]=\cal O$(1), when CP violating effects in the flavor sector are taken into account.
We cure this difficulty by imposing a discrete symmetry and find that this prescription can play additional roles.
It ensures that realistic up-quark mass and Cabibbo angle are simultaneously realized without cancellation between $ \mathcal O(1)$ coefficients.
Also, severe constraints from the chromo electric dipole moment of the quark can be satisfied without destabilizing the weak scale.
The discrete symmetry reduces the number of free parameters, but the model is capable of reproducing quark and lepton mass spectra, mixing angles, and a Jarlskog invariant.
We obtain characteristic predictions $V_{ub} \sim \cal O$($\lambda^4$) ($\lambda=0.22$) and $| V_{cb} Y_b | = |Y_c|$ at the grand unified theory scale.
%

\end{abstract}
\end{titlepage}


\section{Introduction}

Low-energy supersymmetry (SUSY) is one of the leading candidates for physics beyond the standard model (SM) \cite{Nilles:1983ge}.
In particular, the minimal supersymmetric extensions of the SM (MSSM) exhibit remarkable coincidence of three SM gauge coupling constants near $10^{16}$GeV, and the supersymmetric grand unified theory (SUSY GUT) appears to be plausible model  behind the MSSM.
%
%
In these models, however, 
%
if generic soft SUSY breaking terms are introduced, these terms can induce flavor changing neutral current (FCNC) processes and CP violating observables like electric dipole moments (EDMs) that are too large to be consistent with the present experiments \cite{Gabbiani:1996hi}.
These issues are called the SUSY flavor problem and the SUSY CP problem, respectively.
In this paper, we address the later problem in the context of SUSY GUT.
We choose E$_{\text 6}$ as a unification group and introduce an SU(2) flavor symmetry.
Based on this framework, we construct a model of spontaneous CP violation (SCPV) in the flavor sector to provide the origin of the Kobayashi-Maskawa (KM) phase and to evade the SUSY CP problem.
%
%

%
Among several possible grand unification groups \cite{Georgi:1974sy, Fritzsch:1974nn, Gursey:1975ki}, the  E$_{\text 6}$ group with SU(2)$_{\text F}$ flavor symmetry exhibits a numbers of attractive features \cite{Gursey:1975ki,Bando:1999km,Bando:2001bj}.
%
%
%
%
%
\begin{enumerate}
\item
Three fundamental representations of E$_6$, $\bold{ 27}_1$, $\bold{ 27}_2$, and $\bold{ 27}_3$, contain all of the SM fermions and the right-handed neutrinos.
\item
Various realistic hierarchical structures of masses and mixings of quarks and leptons can be naturally derived from the one basic hierarchy of E$_{\text 6}$ invariant Yukawa couplings.
\item 
%
%
%
This basic hierarchy results from the 
SU(2)$_{\text F}$ flavor symmetry and its breaking effects.%
\footnote{
An anomalous U(1)$_\text{A}$ symmetry which we discuss in Sec. 3 also plays a  role to generate hierarchical structures of Yukawa couplings. 
}
\item
The SUSY flavor problem can be avoided by the SU(2)$_{\text F}$ flavor symmetry without destabilizing the weak scale.%
\footnote{
Here we assume that the $D$-term contributions are small.
A possible way to justify this assumption is by adopting non-Abelian discrete symmetry instead of the gauged symmetry which we use in this paper.}
To be more concrete, at the leading order, the model has the following sfermion soft mass matrices at the GUT scale, as a consequence of SU(2)$_{\text F}$ symmetry \cite{Maekawa:2002eh}.
%
%
\begin{equation}
m_{\bold {10}}^2 =
  \left(
    \begin{array}{ccc}
       m_0^2 & 0 & 0 \cr
       0 & m_0^2 & 0 \cr
       0 & 0 & m_3^2
    \end{array}
  \right), \quad
m_{\bar{ \bold {5}}}^2= 
  \left(
    \begin{array}{ccc}
       m_0^2 & 0 & 0 \cr
       0 & m_0^2 & 0 \cr
       0 & 0 & m_0^2
    \end{array}
  \right)
\label{eq:mod-uni}
\end{equation}
(Here, $m_{\bf 10}$ and $m_{\bf\bar 5}$ are the mass matrices of the ${\bf 10}$ and ${\bf\bar 5}$ fields of SU(5), where the SM matter fields are embedded in these multiplets in a standard way \cite{Georgi:1974sy}.)
In this paper, we call this form of sfermion soft mass matrix ``$\it{modified}$ universality"\cite{Maekawa:2002eh, Kim:2006ab, Kim:2008yta, Ishiduki:2009gr, Kim:2009nq}.
In this set up, if the gluino mass ($M_3$), the up-type Higgs mass ($m_{H_U}$), and the supersymmetric Higgs mass ($\mu $) as well as $m_3$ are the weak scale, then the weak scale can be stabilized.%
\footnote{
In order to satisfy the lightest CP-even Higgs boson mass bound set by LEPI$\!$I, one may employ $\it{maximal\, mixing}$ of the stop sector \cite{Carena:1999xa}.
Another interesting option is $\it{light\, Higgs\, scenario}$ or $\it{inverted\, hierarchy}$ where one assumes that the heaviest CP-even Higgs boson behaves as the SM Higgs boson \cite{Kane:2004tk}.
}
Also, even if there are some deviations from 
the ``$\it{modified}$ sfermion universality" in flavor space,
the FCNC constraints can be satisfied by raising $m_0$, and this does not destabilize the weak scale as long as $m_3$ is maintained in the weak scale.\footnote{
However, the muon $g-2$ anomaly is not explained if $m_0 \gtrsim $1TeV, since SUSY contributions decouple \cite{Hagiwara:2006jt}.
}
Note that this prescription is more natural compared to the case for the deviation from 
the ``universality" that is assumed in minimal supergravity  where $m_3$ is set equal to $m_0$.
\end{enumerate}

Now, let us move on to the SUSY CP problem.
%
%
In order to avoid the SUSY CP problem, it is desirable that the dimensionful soft SUSY breaking parameters and the $\mu$ parameter are all real.
On the other hand, the Yukawa couplings should be complex parameters so as to provide a KM phase \cite{Cabibbo:1963yz}.
%
%
%
Is it possible to realize these observations naturally?
To answer this question, in this paper we examine the idea of spontaneous CP violation (SCPV) \cite{Lee:1973iz, Barr:1988wk} within the context of E$_{\text 6}\times $SU(2)$_{\text F}$ SUSY GUT.
First of all, by imposing CP symmetry, all the coupling constants of the original Lagrangian, including soft SUSY breaking terms, can be taken to be real.
Then we make a model in which CP is spontaneously broken by complex vacuum expectation values (VEV) of flavon fields; i.e., the SU(2)$_\text{F}$ flavor symmetry breaking is also responsible for the CP violation.
Here, once renormalizability is assumed, it seems difficult to generate complex Yukawa couplings.
Therefore we allow higher dimensional operators so that complex dimensionless (Yukawa) couplings can be induced via nonrenormalizable interaction terms. 
%
%
Since CP is broken in the SU(2)$_\text{F}$ sector, we can expect that the SU(2)$_\text{F}$ singlet SUSY breaking parameters can remain real after the SCPV, contrary to the Yukawa couplings.
Also, even if flavor nonsinglet SUSY breaking parameters acquire CP phases, their effects on the CP violating observables can be suppressed in the current framework by taking $m_0$ much larger than the weak scale.%
\footnote{
If $m_3$ is placed in the weak scale in view of the naturalness, there are nondecoupling SUSY contributions to the CP violating observables.
These effects can be suppressed if CP is not effectively broken in the up-type (s)quark sector \cite{Ishiduki:2009gr}.
We will comment this point in Secs.4 and 5.
%
%
%
%
}
%
%
%
Therefore, SCPV provides a way to solve the SUSY CP problem.
%

%
%
In the previous papers of Ref.\cite{Maekawa:2001uk}, an anomalous U(1)$_{\text A}$ gauge symmetry \cite{Witten:1984dg, Froggatt:1978nt} was introduced to solve the doublet-triplet splitting (DTS) problem and to provide the origin of the hierarchal structures of Yukawa couplings.
The U(1)$_{\text A}$ symmetry was also applied to solve the $\mu$ problem \cite{Maekawa:2001yh}.
However, as we will see in the following section, if we naively apply the U(1)$_\text{A}$ symmetry to the model of SCPV, it generally leads to an unwanted outcome, Arg$[\mu b^*] =\cal O$(1), and gives rise to the SUSY CP problem.
We cure this difficulty by imposing discrete symmetry.
Interestingly, this discrete symmetry can play additional roles in the model.
It ensures that realistic up-quark mass and Cabibbo angle are  simultaneously realized without cancellation between $ \mathcal O(1)$ coefficients.
Also, the severe constraints from the chromo-EDMs (CEDM) can be satisfied without destabilizing the weak scale.
As a result of the discrete symmetry, the number of free parameters of the model is reduced.
However, the model is capable of reproducing quark and lepton mass spectra, mixing angles, and a Jarlskog invariant.
We also obtain characteristic predictions $V_{ub} \sim \cal O$($\lambda^4$) ($\lambda=0.22$) and $|Y_b V_{cb}| = |Y_c|$ at the GUT scale.
%

%
%
%
%
%
%

This paper is organized as follows.
In the next section, we give a brief review of the model of E$_{\text 6}$ SUSY GUT and SU(2)$_{\text F}$ flavor symmetry.
In Sec.3, an anomalous U(1)$_{\text A}$ gauge symmetry and its specific vacua are summarized.
Based on these two sections, we show a simple model of the SCPV that is caused by the VEVs of flavon fields, in Sec.4.
In the same section, we also discuss CP violation effects on $\mu$- and $b$-terms and indicate that they lead to Arg$[\mu b^*] =\cal O$(1).
Then, we introduce discrete symmetry in order to  solve this difficulty.
In Sec.5, we summarize a model of SCPV in E$_{\text 6}$ SUSY GUT with SU(2)$_\text{F} $, U(1)$_\text{A}$, and discrete symmetries.
Then, we examine the consequent Yukawa couplings and derive some predictions of the model.
The last section is devoted to a summary and discussion.

\section{E$_{\text 6}\times$SU(2)$_{\text F}$ SUSY GUT}
\label{sec:e6*su2 susy gut}

In this section we briefly summarize the model of E$_{\text 6}\times$SU(2)$_{\text F}$ SUSY GUT discussed in Refs.\cite{Bando:2001bj, Maekawa:2002eh}.
We mainly focus on the Yukawa sector and sfermion soft masses of the model.

\subsection{E$_6$ GUT}
\label{subsec:e6 gut}

In this subsection, we discuss the Yukawa sector of E$_{\text 6}$ SUSY GUT.
Here the twisting mechanism among SU(5) ${\bar {\bold{ 5 }}}$ fields plays important roles \cite{Bando:1999km, Bando:2001bj, Maekawa:2002eh}.

For the E$_{\text 6}$ group, $\textbf{27}$ is the fundamental representation, and in terms of E$_{ 6}\supset $SO(10)$\times $U(1)$_{V'}$ (and [SO(10)$\supset$SU(5)$\times$U(1)$_{V}$]) it is decomposed as%
\footnote{
Here acutes are used to distinguish different $\bar {\textbf{5}}$($\textbf{1}$)s.
Each subscript represents corresponding U(1) charge.
}
\begin{eqnarray}
\textbf{27} = \textbf{16}_1 [ \textbf{10}_{-1} + \bar {\textbf{5}}_3 + \textbf{1}_{-5}] 
+ \textbf{10}_{-2} [ \textbf{5}_2 + \bar{\textbf{5}}_{-2}^\prime ] + \textbf{1}_4 [ \textbf{1}_0^\prime]
 \, .
\label{eq:e6 decomposition}
\end{eqnarray}
As shown in \eqref{eq:e6 decomposition}, each $\textbf{27}$ incorporates two $\bar {\textbf{5}} $'s(and ${\textbf{1}} $'s) of SU(5).
This nature allows a model to simply produce different and realistic Yukawa structures of quark and lepton from a single hierarchical structure of an E$_6$ invariant Yukawa coupling \cite{Bando:1999km, Bando:2001bj}.

Let us introduce the following superpotential:
\begin{eqnarray}
W_{\textnormal E_{ 6}} \supset
Y_{ij}^H \Psi_i^{\textbf{27}} \Psi_j^{\textbf{27}} H^{\textbf{27}} + Y_{ij}^C \Psi_i^{\textbf{27}} \Psi_j^{\textbf{27}} C^{\textbf{27}}
\label{eq:E_6 Yukawa}
\end{eqnarray}
and assume that the original Yukawa hierarchies are
\begin{eqnarray}
Y_{ij}^{H} \sim Y_{ij}^{C} \sim \left( 
\begin{array}{c c c}
\lambda ^6 & \lambda ^5 & \lambda ^3 \\
\lambda ^5 & \lambda ^4 & \lambda ^2 \\
\lambda ^3 & \lambda ^2 & 1 
\end{array}
\right) .
\label{eq:original Yukawa}
\end{eqnarray}
Here $\Psi_i^{\textbf{27}} \, (i=1-3) $ are matter fields, and $H ^{\textbf{27}} $ and $C ^{\textbf{27}}$ are Higgs fields that break E$_{ 6}$ into SO(10) and SO(10) into SU(5), respectively.\footnote{In the following discussion, we also introduce antifundamental fields, ${\bar H}^{\bar {\bold{ 27 }}}$ and ${\bar C}^{\bar {\bold{ 27 }}}$, to maintain corresponding D-flatness conditions.
Also, the adjoint field $A ^{\textbf{78}} $ is introduced to realize the DTS, employing the Dimopoulos-Wilczek (DW) mechanism \cite{Dimopoulos:1981xm}.
A complete list of field contents is given in Sec.\ref{sec:a modle of scpv in e6 susy gut}.
}
We assume that the MSSM Higgs doublets are included in $H ^{\textbf{27}} $ and $C ^{\textbf{27}}$.
Hereafter, we parametrize the several hierarchical structures of couplings and VEVs of various fields in terms of the Cabibbo angle $\lambda$.
(We set $\lambda \equiv 0.22$.)
In Eq.\eqref{eq:original Yukawa}, we only parametrize the hierarchical structures, and the $\cal O$(1) coefficients are suppressed for the moment.

In Eq.\eqref{eq:e6 decomposition}, once $H^{\textbf{27}} $ and $C^{\textbf{27}}$ acquire VEVs in the components of $\textbf{1}_4 ( \textbf{1}^\prime_0 )$ and $\textbf{16}_1 (\textbf{1}_{-5})$, respectively, a superheavy mass matrix of rank 3 among $\bold{ 5}_i, \bar {\bold{ 5 }}^\prime_i$, and $\bar {\bold{ 5 }}_i$ of $\Psi_i^{\textbf{27}}$ is induced, through the Yukawa coupling of Eq.\eqref{eq:E_6 Yukawa}.
Therefore the 3 degrees of freedom among $\bar {\bold{ 5 }}^\prime_i$ and $\bar {\bold{ 5 }}_i$ decouple at the GUT scale.
In consequence, the up-type quark Yukawa coupling ($Y_U$) remains in the original form of Eq.\eqref{eq:original Yukawa} but  the down-type quark ($Y_D$) and charged lepton Yukawa couplings ($Y_E$) differ from it.
%
%
Importantly, the three massless modes among $\bar {\bold{ 5 }}^\prime_i$ and $\bar {\bold{ 5 }}_i$ mainly originate from the first two generations of $\Psi_{i}^{27} $ because the third generation $\bar {\bold{ 5 }}^\prime_3$ and $\bar {\bold{ 5 }}_3$ from $\Psi_{3}^{27} $ have large Yukawa couplings, i.e., large GUT scale masses.
%
%
This feature is also important for making sfermion universality for the ${\bar {\bold{ 5 }}}_{}$ sector, as discussed in the following subsection.
%
As an example, when $\langle C \rangle/ \langle H \rangle \sim \lambda^{0.5}$, we end up with the following milder hierarchies for $Y_D$ and $Y_E$ \cite{Bando:2001bj}:
%
%
\begin{eqnarray}
Y_D \sim Y_E^T \sim \left( 
\begin{array}{c c c}
\lambda ^6 & \lambda ^{5.5} &\lambda ^5 \\
\lambda ^5 & \lambda ^{4.5} &\lambda ^4 \\
\lambda ^3 & \lambda ^{2.5} &\lambda ^2 
\end{array}
\right).
\label{eq:yd and ye}
\end{eqnarray}
%
%
Note that the hierarchies of Eqs.\eqref{eq:original Yukawa} and \eqref{eq:yd and ye} are adequate to reproduce several mass spectra and  flavor mixing angles of quarks and leptons.%
\footnote{See Ref.\cite{Bando:2001bj} for details.}
%
%
However, an exception is the  up-quark Yukawa coupling $Y_u$.
The hierarchy of Eq.\eqref{eq:original Yukawa} leads to $Y_u \sim \lambda^6$ but the realistic value is $Y_u \sim \lambda^8 $.\footnote{See, e.g., Ref.\cite{Fusaoka:1998vc, Ross:2007az}.}
%
%
We will get back to this issue in Sec.\ref{subsec:susy cp problem and discrete symmetry}

%

\subsection{SU(2) flavor symmetry and modified universality}
\label{subsec:su(2)f and modified universality}

In this subsection, we briefly discuss the effect of SU(2)$_{\text F}$ flavor symmetry on the sfermion soft masses and see the emergence of the $\it{modified}$ universal form of Eq.\eqref{eq:mod-uni} \cite{Maekawa:2002eh}.

First of all, we assume that the original hierarchy of Eq.\eqref{eq:original Yukawa} (partly) originates from the flavor symmetry breaking effects.
In order to naturally incorporate the $\cal O$(1) top quark Yukawa coupling, here, we employ SU(2)$_{\text F}$ flavor symmetry and treat the first two generations of matter fields as the doublet $\Psi^{\textbf{27}}_a$ whereas the third generation field $\Psi_3^{\textbf{27}}$ and all the Higgs fields are treated as singlets under SU(2)$_{\text F}$.
We also introduce flavon fields $F_a$ and $\bar F^a$ that are singlets under E$_{\text 6}$ and a doublet and an anti-doublet under SU(2)$_{\text F}$, respectively.
Then, as shown later, $F_a$ and $\bar F^a$  acquire VEVs and break the the SU(2)$_{\text F}$.
These effects generate hierarchical structures of the Yukawa coupling of Eq.\eqref{eq:original Yukawa}.


Suppose that the soft SUSY breaking terms are mediated to the visible sector above the scale where E$_{\text 6}$ and SU(2)$_{\text F}$ symmetries are respected, such as in gravity mediation.
Then, $\Psi^{\textbf{27}}_a$ acquires a soft mass ($m_0$) that is different from the mass of $\Psi^{\textbf{27}}_3$ ($m_3$), in general, in an E$_{\text 6}\times$SU(2)$_{\text F}$ symmetric way.
As a result, this guarantees whole sfermion soft mass degeneracy at the GUT scale, except for $\textbf{10}_3$, as shown in Eq.\eqref{eq:mod-uni},
%
%
because ${\bold{ 10 }_1}$, ${\bold{ 10 }_2}$, 
and all the MSSM ${\bar {\bold{ 5 }}}$ fields originate from an identical field $\Psi^{\textbf{27}}_a$.\footnote{
Since possible deviations from Eq.\eqref{subsec:su(2)f and modified universality}, which are induced by the SU(2)$_{\text F}$ and E$_6$ breaking effects, are argued in Ref.\cite{Maekawa:2002eh}, we do not repeat the detailed argument here but the FCNC constraints can be satisfied in the current model by employing the argument given in the Introduction.
}

\section{Anomalous U(1) gauge symmetry}
\label{sec:u1a}

In the previous papers of Ref.\cite{Maekawa:2001uk}, an anomalous U(1)$_{\text A}$ gauge symmetry \cite{Witten:1984dg, Froggatt:1978nt} was introduced to solve the doublet-triplet mass splitting and to provide the origin of the hierarchical structures of Yukawa couplings.
In this section, we summarize essential points and comment on what kinds of fields acquire VEVs, and how the magnitudes of these VEVs are determined in the framework of U(1)$_{\text A}$.

We introduce an anomalous U(1)$_{\text A}$ gauge symmetry whose anomalies are canceled by the Green-Schwarz mechanism \cite{Green:1984sg}.
The theory possesses the Fayet-Iliopoulos term ($\xi^2$), and we assume its magnitude as $\xi = \lambda \Lambda$.
(Here $\Lambda$ is the cutoff scale of the theory and we set $\Lambda=1$.)
Let us denote the symmetries of the theory, except for U(1)$_{\text A}$, as $G_a $.
Then it is shown in Ref.\cite{Maekawa:2001uk} that a system consisting of all the $G_a$ and U(1)$_{\text A}$ invariant terms\footnote
{
Here we include all the nonrenormalizable operators.
Also, each term is assumed to have an $\cal O$(1) coefficient.
}
 has a supersymmetric vacuum where all the fields that are negatively charged under U(1)$_{\text A}$ get VEVs in the following way.\footnote
{
From now on, each superfield is denoted by an uppercase letter, whereas the corresponding lower case letter indicates an associated U(1)$_{\text A}$ charge.
The consistency of Eq.\eqref{eq:u1a vev} requires the number of positively charged fields to be larger than that of negatively charged fields by one.	
}
\begin{eqnarray}
\left \{
\begin{array}{lc}
\langle Z_i^+ \rangle =  0 \nonumber & (z_i^+>0) \\
\langle Z^-_i \rangle \simeq \lambda^{-z^-_i} & (z_i^-<0)
\end{array}
\right. \quad \eqref{eq:u1a vev}
\label{eq:u1a vev}
\end{eqnarray} 
Here $Z_i$ denotes the $G_a$ singlet field, but the argument can be extended to the case where $Z_i $ are composite operators that are made by $G_a $ nonsinglet fields.
For example, in case the where $Z^-= \bar X X $ (here $X$ denotes  the $G_a $ nonsinglet field and $\bar X$ is the antirepresentation of $X$),  $\langle \bar X X \rangle = \lambda^{-(x + \bar x)} $ leads to $\langle X \rangle = \langle \bar X \rangle = \lambda^{-(x + \bar x)/2}  $, once the D-flatness condition of $G_a $ is taken into account.

A U(1)$_{\text A}$ symmetry and its specific SUSY vacuum are applied in several aspects of phenomenological model building.
For example, an appropriate U(1)$_{\text A}$ charge assignment for the Higgs sector can ensure the DTS via the DW mechanism \cite{Maekawa:2001uk, Dimopoulos:1981xm}.
Here note that since a positively charged field does not acquire a VEV in this U(1)$_{\text A}$ vacuum, only the terms that are linear in a positively charged field are relevant to the determination of the symmetry breaking structure of the model.
%

It is also important to mention that a term whose total U(1)$_{\text A}$ charges are negative does not appear at the U(1)$_{\text A}$ breaking vacuum.
The reason is that this type of term should originally be accompanied by at least one positively charged field but its VEV is always vanishing according to Eq.\eqref{eq:u1a vev} (SUSY-zero mechanism) \cite{Nir:1993mx, Maekawa:2001uk}.
Importantly, this property can be applied to solve the $\mu$ problem \cite{Maekawa:2001yh}.
To be more precise, when the U(1)$_{\text A}$ charges of the Higgs fields are set to be negative, the supersymmetric $\mu$-term is forbidden by the SUSY-zero mechanism.
However, this SUSY-zero mechanism will be broken by the amount of the SUSY breaking scale, once soft SUSY breaking effects are taken into account in the total Lagrangian.
So the appropriate scale of the $\mu$-term can be induced.
Employing these properties of U(1)$_{\text A}$, in the following sections we construct and examine a model of SCPV in the E$_{\text 6}$ SUSY GUT.
%
%
%
%
%
%
%

\section{Spontaneous CP violation}
\label{sec:spontaneous cp violation}

In this section, 
%
we first discuss the mechanism of SCPV  that is caused by the SU(2)$_{\text F}$ breaking sector.
After specifying a source for the CP violating phase, we examine how this phase will appear in the parameters of the MSSM.
Particularly, we refer to a solution for the $\mu$ problem in the context of U(1)$_{\text A}$ and indicate that it leads to an unwanted outcome, Arg$[\mu b^*] = \cal O$(1), once CP violating effects are taken into account.
Next, we discuss discrete symmetry in order to solve this difficulty.
%

\subsection{Spontaneous CP violation}
\label{subsec:spontaneous CP violation}

In this subsection, adopting a U(1)$_{\text A}$ symmetry and its characteristic vacuum, we discuss the mechanism of SCPV that is caused by the SU(2)$_{\text F} $ breaking sector.

Let us introduce the E$_{\text 6} \times $SU(2)$_{\text F}$ singlet field $S (s > 0)$.
Then, in general, we obtain the following superpotential  made of $F_a (f<-1) $, $\bar F^a ( \bar f <-1)$, and $S$:
\begin{eqnarray}
W_S = \lambda^{s}S [ \sum_{n=0}^{n_f } c_{n} \lambda^{ (f+\bar f) n } (\bar F F)^{n} ]
\label{eq:general Ws}
\end{eqnarray}
%
%
We assume that $\cal O$(1) coefficients $c_{n}$ are all real and that the theory is originally CP invariant.\footnote{
In general, following terms can also appear in the bracket of Eq.\eqref{eq:general Ws}.
\begin{eqnarray}
\Sigma \lambda^{(h+ \bar h) n_h} (H^
\bold{ 27 } {\bar H}^{\overline {\bold{ 27 }}})^{n_h} + 
\dots 
\label{eq:possible-contribution-to-Ws}
\end{eqnarray}
However, these terms are not important for the determination of $\langle \bar F F \rangle$.
From Eq.\eqref{eq:u1a vev}, it is understood that their signatures are summarized in $c_0$ of Eq.\eqref{eq:general Ws}.
}
$n$ are natural numbers which are truncated at $n_f$, where $s+ (f+\bar f) (n_f + 1) $ becomes negative (SUSY-zero mechanism).
The appearance of $\lambda $ in Eq.\eqref{eq:general Ws} can be understood as originating from the VEV of an operator whose U(1)$_{\text A}$ charge is $-1$.
From Eq.\eqref{eq:general Ws}, the $F$-flatness condition with respect to $S$ leads to
\begin{equation}
\lambda ^s[c_0 + c_1 \lambda ^{(f+\bar f)} \langle \bar FF \rangle +\dots +c_{n_f} \lambda  ^{n_f (f+\bar f)}  \langle \bar FF \rangle ^{n_f}]=0 \, .
\label{F=0}
\end{equation}
Thus, when $n_f \geq 2$, it generally forces $\langle \bar F F \rangle $ to acquire an imaginary phase.
Therefore, at this stage the CP invariance is spontaneously broken.
%
Using SU(2)$_{\text F}$ gauge symmetry and taking its $D$-flatness condition into account, hereafter we adopt the following basis where only $\langle F_2 \rangle $ acquires an imaginary phase:
\begin{eqnarray}
\langle F_a \rangle
\sim 
\left( 
\begin{array}{c}
0 \\
 e^{i\rho }\lambda ^{-\frac{(f+\bar f)}{2}}  
\end{array}
\right),
\indent
\langle \bar F^a \rangle
\sim 
\left( 
\begin{array}{c}
0 \\
\lambda ^{-\frac{(f+\bar f)}{2}}  
\end{array}
\right)
\label{<F_a>}
\end{eqnarray}

\subsection{Spontaneous CP violation and $\mu$-term generation}
\label{subsec:spontaneous CP violation and mu-term generation}

In this subsection, 
%
%
we discuss the effect of Eq\eqref{<F_a>} on the generation of the $\mu$- and $b$-terms.
Here an appropriate $\mu$-term is induced by the SUSY breaking effect on the SUSY-zero mechanism.%
\footnote{
Note that, within the context of the U(1)$_{\text A}$  summarized in the previous section, the Giudice-Masiero mechanism \cite{Giudice:1988yz} does not work well since it requires $\abs{2 h} \leq 1$ \cite{Nir:1995bu}.
}
%
%

%

%
Let us assume that the MSSM Higgs doublets are contained in $\bold{ 5 }$ and ${\bar {\bold{ 5 }}'}$ of $H^\bold{27}$.
Then if the U(1)$_{\text A}$ charge of $H^\bold{27}$ is negative (h$<$0), there is no supersymmetric mass term for the MSSM Higgs doublets.
This is because the positively charged field $S$ does not acquire VEVs in the SUSY limit (SUSY-zero mechanism).
Here, a possible term is $\lambda^{s+3h} S H^{\bold{ 27 }}H^{\bold{ 27 }}H^{\bold{ 27 }} \rightarrow  \lambda^{s+3h}  \langle S \rangle \langle H^{\bold{ 1 }} \rangle H^{\bold{ 10 }}H^{\bold{ 10 }}$, but $\langle S \rangle $ is vanishing in the SUSY limit.

\begin{figure}[]
\begin{center}
\includegraphics[width=.4\textwidth]{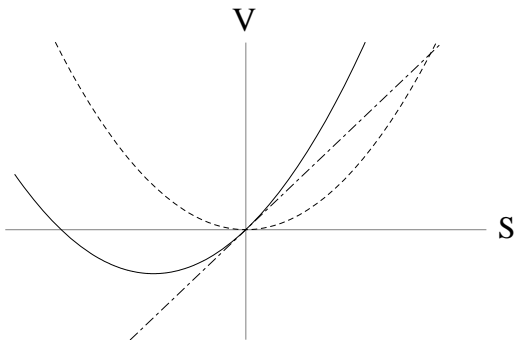}
\caption{
SUSY breaking effect on the potential of $S$
{\footnotesize}
}
\label{SUSY breaking effect}
\end{center}
\end{figure}

However, the situation changes once we take SUSY breaking effects into account in the total Lagrangian \cite{Hall:1983iz, Maekawa:2001yh}.
For a schematic explanation, we first discuss a simplified version of the system which consists of E$_{\text 6}\times$SU(2)$_{\text F}$ singlet fields $S$($s>0$) and $Z$($z<-1$).
%
The superpotential is given as
\begin{eqnarray}
W \supset \lambda^{s}S + \lambda^{s+z}SZ + \dots 
\label{eq:superpotentials for s}
\end{eqnarray}
and we also introduce the corresponding general soft SUSY breaking terms%
\footnote{
In Eq.\eqref{eq:superpotentials for s}, abbreviation denotes $\lambda^{2s+z}S^2Z$,
$\lambda^{s+2z}SZ^2$, etc.
Since SUSY breaking effects does not largely change Eq(3.1), these terms are not relevant and omitted in the following arguments.
This is the same for Eq.\eqref{eq:soft-terms-sz}.}
\begin{eqnarray}
V_{soft} = \textstyle \sum \limits_{X=S,Z} \, \tilde m_X^2 |X|^2 + A_{S} \lambda^{s}S + A_{Z} \lambda^{s+z}SZ + \dots .
\label{eq:soft-terms-sz}
\end{eqnarray}
Here we assume that all the soft parameters $\tilde m_S$, $\tilde m_Z$, $A_S $ and $A_Z $ are the weak scale.
As usual, the whole potential is composed of $V= \textstyle \sum \limits _{X=S,Z} \, |F_X|^2 + V_{soft} \, $,%
\footnote{
Note that the U(1)$_\text{A}$ D-flatness is satisfied by the VEV of an operator whose U(1)$_{\text A}$ charge is $-1$.
}
where
\begin{eqnarray}
|F_S|^2 = |\frac{\delta W}{\delta S}|^2 = | \lambda^{s} + \lambda^{s+z}  Z |^2, \quad
|F_Z|^2 =|\frac{\delta W}{\delta Z}|^2 = |\lambda^{s+z}S |^2.
\label{eq:fz}
\end{eqnarray}
Here note that, in the SUSY limit, $\langle Z \rangle$ is determined so that $|F_S|$ of Eq.\eqref{eq:fz} vanishes.
Also, $\langle S \rangle =0$ satisfies the vanishing $|F_Z|$.
However, once SUSY breaking effects are taken into account, there is no need for $F$-flatness conditions to be satisfied.
Actually, the total potential $V$ which includes SUSY breaking effects makes $\langle S \rangle $ and $\langle F_S \rangle $ nonvanishing.
First of all, the extremum condition of $V$ with respect to $S$ determines $\langle S \rangle$ as%
\begin{eqnarray}
\langle S \rangle \sim 
\lambda^{-(s+2z)} (A_S+A_Z).
\label{eq:<S>}
\end{eqnarray} 
Here the shift of $\langle S \rangle $ from zero can be understood as the balance between the SUSY mass term and the tadpole term that is induced by the SUSY breaking effect.
\footnote{We plot the potential respect to $S$ in Fig.\ref{SUSY breaking effect}.
Here the dotted parabola shows the SUSY mass term ($|\lambda^{s+z}S|^2$) and the dashed-dotted line indicates the SUSY breaking tadpole contribution ($A_S \lambda^{s} S + A_Z \lambda^{s+z} S\langle Z \rangle$).
The solid line is the sum of these contributions.
}
Also, $F_S$ is fixed by the extremum condition of V with respect to $\Delta Z$, where $\Delta Z$ is defined as the deviation of $\langle Z \rangle$ from its SUSY VEV as $\langle Z \rangle =\lambda^{-z}+\Delta Z$.
\begin{eqnarray}
\frac{\delta V}{\delta \Delta Z} &=& F_S^* \lambda^{s+z} + \tilde m_Z^2 \lambda^{-z} + A_Z \lambda^{s+z} S=0 \nonumber \\
&\rightarrow &
\langle F_S^* \rangle= \lambda^{-(s+2z)} \tilde m_Z^2 +A_Z \langle S \rangle 
\label{eq:<F_s>}
\end{eqnarray}
Combining Eqs.\eqref{eq:<S>} and \eqref{eq:<F_s>},  the $\mu$- and b-terms are induced as follows:
%
\begin{eqnarray}
{\cal L} &\supset& \lambda^{s+3h} S H H \langle H \rangle |_{\theta ^2}+ 
\lambda^{s+3h} A_{SH}S H H \langle H\rangle \nonumber \\
&\rightarrow& \lambda^{s+ 2h + \frac{h - \bar h}{2}} ( \langle S\rangle HH |_{\theta ^2} +  (\langle F_S \rangle+\langle S\rangle A_{SH}) HH ) 
\label{eq:myu1}
\end{eqnarray}
Therefore, when $z$ is appropriately chosen, the $\mu$ and $b$ parameters appear as the weak scale.

Now, let us examine the effects of the CP violation of Eq.\eqref{<F_a>} on the generation of the $\mu$ and $b$ parameters.
So far we have assumed that all the soft SUSY breaking terms and $\cal O$(1) coefficients were real.
Therefore, the $\mu$ and $b$ parameters are real.
However, this result can be changed, in general, once flavon fields acquire complex VEVs.
In order to see this, we extend Eq.\eqref{eq:superpotentials for s} to include Eq.\eqref{eq:general Ws}.
Also, we take the corresponding SUSY breaking terms
\begin{eqnarray}
A \lambda^{s} S [ \sum_{n=0}^{n_f } c'_{n} \lambda^{ (f+\bar f) n } (\bar F F)^{n} ]
\label{eq:complex A}
\end{eqnarray}
into account.
Here note that each $\cal O$(1) coefficient $c_n'$ appearing in Eq.\eqref{eq:complex A} is different from that of Eq.\eqref{eq:general Ws}, in general.%
\footnote
{
When each $ \mathcal O(1)$ coefficient appearing in a superpotential term equals to the corresponding coefficient of a soft SUSY breaking term, the above mechanism for the generation of $\mu$-term does not work.
}
Therefore,
\begin{eqnarray}
A \lambda^{s} S \{ c_0' + c_1' \lambda^{ f + \bar f} \langle \bar F F \rangle + \dots + \Sigma c_{n_f}' \lambda^{ (f + \bar f) n_f} \langle \bar F F \rangle ^{n_f} \}
\label{eq:non-vanishing A}
\end{eqnarray}
is generically nonvanishing, even though the left-hand side of \eqref{F=0} vanishes.
%
Note that, from Eq.\eqref{eq:u1a vev},  each higher dimensional operator in Eq.\eqref{eq:non-vanishing A} gives a contribution that is of the same order of the original tadpole term, $c'_0A \lambda^s S$, but the effective coefficient is complex.
Therefore, Eq.\eqref{eq:complex A} turns into $A_S \lambda^s S$ of Eq.\eqref{eq:soft-terms-sz}, where $A_S$ is complex and Arg$[A_S] \sim \mathcal O(1)$.
Then, we conclude from Eqs.\eqref{eq:<S>}, \eqref{eq:<F_s>} and \eqref{eq:myu1} that the $\mu$ and $b$ parameters turn out to be complex parameters and Arg$[ \mu b^*] = \cal O$(1), in general.

\subsection{SUSY CP problem and discrete symmetry}
\label{subsec:susy cp problem and discrete symmetry}

As we saw in the previous subsection, CP violating VEVs of flavon fields generically lead to complex $\mu$ and $b$ parameters and Arg$[ \mu b^*] = \cal O$(1).
This rephasing invariant complex phase induces (C)EDMs of quarks and leptons, and gives rise to the SUSY CP problem.
Though the decoupling procedure can reduce the SUSY contribution to the (C)EDMs, we discuss another way to avoid this issue in order not to destabilize the weak scale.


As we discussed below Eq.\eqref{eq:complex A}, the couplings among $S$ and flavon fields led to complex $\mu$ and $b$ parameters and Arg$[ \mu b^*] = \cal O$(1).
Therefore, we introduce a discrete symmetry so that coupling among $S$ and $\bar F F$ is forbidden.%
\footnote{
In actual model building we should introduce an additional positive singlet field $S'$ which plays a role of $S$ appeared in Eq.\eqref{eq:general Ws}.
Then this $S'$ may acquire complex VEV in its scalar and F components.
However, the coupling between $S'$ and $H^3$ can be forbidden by an appropriate symmetry.
Or, since the magnitude of $\langle S' \rangle $ is dictated by the smallest negative charge of the relevant field, another $Z'$ field can make $\langle S' \rangle $ sufficiently small, and ensures $\mu$- and $b$-terms to be almost real.
}
Specifically, we exploit the discrete symmetry in order to discriminate $F$ from the others assigning a nontrivial discrete charge for $F$, since only $F$ acquires a complex VEV in the basis defined in Eq.\eqref{<F_a>}.
One may think that the introduction of an additional symmetry seems ad hoc, but we can find the utility of this discrete symmetry from several different point of view, as follows.

As we mentioned at the end of Sec.2.1, a naive assumption for the Yukawa hierarchy of Eq.\eqref{eq:original Yukawa} leads to a  relatively large up-quark Yukawa coupling.
Therefore, there should be cancellation between $\cal O$(1) coefficients of Eq.\eqref{eq:original Yukawa}, in order to reproduce realistic up-quark Yukawa coupling ($Y_u \sim \lambda^8$) \cite{Fusaoka:1998vc, Ross:2007az}.
However, the situation can be improved if the following types of interactions are only responsible for the generation of $Y_H$:
\begin{eqnarray}
\left( 
\begin{array}{ccc}	
0   & \lambda ^{2 \psi_a + a} \Psi^{a \bold{27}} A^\bold{78} \Psi_a^\bold{27} & 0 \\
\lambda ^{2 \psi_a + a} \Psi^{a \bold{27}} A^\bold{78} \Psi_a^\bold{27} & \lambda ^{2( \psi_a + \bar f)} \bar F^a \Psi_a^\bold{27} \bar F^b \Psi_b^\bold{27} & \lambda^{\psi_a + \psi_3 + \bar f} \bar F^a \Psi_a \Psi_3 \\
0 & \lambda^{\psi_a + \psi_3 + \bar f} \Psi_3 \bar F^a \Psi_a & \lambda^{2 \psi_3} \Psi_3 \Psi_3
\end{array}
\right) \lambda ^h H^\bold{27} \nonumber \\
\rightarrow
\left( 
\begin{array}{ccc}
0   & Q_{B-L} \lambda^5 \Psi_1 \Psi_2  & 0\\
Q_{B-L} \lambda^5 \Psi_2 \Psi_1 & \lambda^4 \Psi_2 \Psi_2 & \lambda^2 \Psi_2 \Psi_3 \\
0 & \lambda^2 \Psi_3 \Psi_2 & \Psi_3 \Psi_3
\end{array}
\right) H \label{eq:yh110}
\end{eqnarray}
%
%
%
Here $A^{\bold{78}}$ is assumed to acquire $B-L$ conserving VEVs \cite{Maekawa:2001uk} and $Q_{B-L}$ represents the $B-L$ charge of the corresponding component of $\Psi^{\bold{27}}$.
Then, Eq.\eqref{eq:yh110} leads to $Y_u \sim Q_{B-L}^2  \lambda^6 $, which can be an order of magnitude smaller than the original expectation, $Y_u \sim   \lambda^6 $.%
\footnote{
In general $Y_U$ and $Y_D$ are the combinations of $Y_H$ and $Y_C$.
However, we make Higgs sector so that the MSSM up-type Higgs doublet  $H_U$ originates solely from $H_u [H^\bold{ 5 }$] but the down-type Higgs doublet $H_D$ originates as $H_D= H_d[H^{{\bar {\bold{ 5 }}'}}$]+$\lambda^{0.5}  L $[$C^{{\bar {\bold{ 5 }}}}$].
(See Appendix \ref{sec:higgs sector}.)
In this case, $Y_D$ is combinations of $Y_H$ and $Y_C$ but $Y_U$ is directly given by $Y_H$.
\label{footnote:Higgs mixing}
}
Simply, Eq.\eqref{eq:yh110} is ensured if there is a discrete symmetry which discriminates $F$ from the others and forbids $ F^a \Psi_a^\bold{27} F^b \Psi_b^\bold{27} H^\bold{27} $ and $ F^a \Psi_a^\bold{27} \bar F^b \Psi_b^\bold{27} H^\bold{27} $.%
\footnote{
Note that, in this case, $F^a \Psi_a \Psi_3 \rightarrow \lambda^3 \Psi_{1} \Psi_{3}$ is also forbidden.}

Also, another utility of discrete symmetry arise from the examination of the SUSY CP problem that is caused even if the $\mu$- and $b$-terms are real.
In the case where the $\it{modified}$ sfermion universality of Eq.\eqref{eq:mod-uni} is adopted, the nonuniversality of the up-type squark sector ($m_{\bold{10}}^2$) induces quark a CEDM \cite{Hisano:2004tf, Ishiduki:2009gr}.
This CEDM exceeds experimental bound if $m_3$ is placed in the weak scale.
%
Here, the CP violating phase is provided by the up-type quark Yukawa coupling, even if $\mu$, $b$, $M_3$, and $A$ parameters are real.
Therefore, when we respect the weak scale stability, the  up-type quark Yukawa coupling should also be real parameters other than $M_3$, $\mu$, and $b$.%
\footnote{
We assume that each Yukawa coupling has a corresponding interaction with the SUSY breaking spurion field ($X=\theta^2 m_{SUSY}$), like $c_{ij}^{H(C)} X {Y}_{ij} \Psi_i \Psi_j H(C)$, where $c_{ij}^{H(C)}$ are the real $ \mathcal O(1)$ coefficient.
Then if $Y_f$ is real (complex), the corresponding $A_f$ term is also real (complex).
}
This requirement can also be ensured if there is a discrete symmetry which discriminates $F$ from the others and $F$ does not contribute to the generation of $Y_H$, since only $F$ acquires complex VEV.%
\footnote{
See \ref{footnote:Higgs mixing}.
}%

In connection with this type of SUSY CP problem that is caused by the flavor off-diagonal complex soft term, the SU(2)$_\text{F}$ breaking effect can also induce similar flavor off-diagonal complex soft term.
For example, a higher dimensional term in the K$\ddot {\text a}$hler potential, $\tilde m^2 (\Psi_3^\bold{27})^\dagger \epsilon^{ab} F_a \Psi_b^\bold{27}$, 
induces a complex $(m_{\bf 10}^2)_{13}$ 
component of the order of $e^{i \rho} \lambda^{3} \tilde m^2 $, and it contributes to the up-quark (C)EDM.
However, this effect can also be forbidden if there is a discrete symmetry which discriminates $F$ from the others.

Following these observations, we introduce a discrete symmetry to our model so that all the above requirements can be fulfilled.
When we adopt a cyclic symmetry, $Z_3$ is the smallest group that can allow $ \bar F^a \Psi_a^\bold{27} \bar F^b \Psi_b^\bold{27} H^\bold{27} $ but forbid $ F^a \Psi_a^\bold{27} F^b \Psi_b^\bold{27} H^\bold{27} $.
%
Then, e.g., $F$  is assigned $Z_3=1$ and all the other fields are assigned $Z_3=0$, in order to discriminate $F$ from the others.
Here, an exception is $C^\bold{27} $.
As explained in the next section, the model can not be realistic without $F^a \Psi_a^\bold{27} \bar F^b \Psi_b^\bold{27} C^\bold{27} $ and $F^a \Psi_a^\bold{27} \Psi_3^\bold{27} C^\bold{27} $, since it supplies the KM phase through the down-type quark Yukawa coupling.
Therefore, $C^\bold{27}$ should be assigned $Z_3=2$.
Once a Higgs field is charged under this new $Z_3$ symmetry, some modifications are necessary in the Higgs sector that is originally composed to ensure the DTS \cite{Maekawa:2001uk}.
In the following discussion, therefore, we extend $Z_3$ into $Z_6$ where $Z_6$ incorporates the $Z_2$ symmetry that was used to ensure that the adjoint Higgs field acquired B-L conserving VEV \cite{Barr:1997hq, Maekawa:2001uk}.
Here, a simple choice of $Z_6$ charges is $Z_6[F]=2$ and $Z_6[C]=4$. However, we set $Z_6[F]=1$ and $Z_6[C]=5$ in order to utilize the degrees of freedom of the enlarged discrete symmetry.
%


%
%

%

\section{A model of SCPV in E$_{\text 6}$ SUSY GUT}
\label{sec:a modle of scpv in e6 susy gut}

In this section, we summarize a model of SCPV in the E$_6$ SUSY GUT with SU(2)$_{\text F}$, U(1)$_{\text A}$, and Z$_{\text 6}$ symmetries.
After giving brief explanations for the field contents, we focus on the effects of the discrete symmetry on the Yukawa couplings of the model.
We examine consequent Yukawa couplings and derive some predictions from them.

\subsection{Field content}
\label{subsec:field contens}

Here we summarize the field content of the model and its representations under the E$_{\text 6}$, SU(2)$_{\text F} $, U(1)$_{\text A}$ and $Z_6$ symmetries.

\begin{table}[]
\begin{center}
\begin{tabular}{cccccccccccccccc}
\toprule
 &$\Psi_a^\bold{27}$&$\Psi_3^\bold{27}$&$F_a$&$\bar F^a$&$H^\bold{27}$&$\bar H^{\overline {\bold{27}}} $&$C^{\bold{27}}$&$\bar C^{\overline {\bold{27}}}$&$C'^{\bold{27}}$&$\bar C'^{\overline {\bold{27}}}$&$A^{\bold{78}}$&$A'^{\bold{78}}$\\
\midrule
$E_6$ & $\bold{27}$    & $\bold{27}$  & $\bold{1}$ & $\bold{1}$ & $\bold{27}$&${\overline {\bold{27}}}$&$\bold{27}$&${\overline {\bold{27}}}$&$\bold{27}$&${\overline {\bold{27}}}$&$\bold{78}$ &$\bold{78}$ \\
$SU(2)_F$& $\bold{2}$ &$\bold{1}$&$\bold{2}$&$\bar{\bold{2}}$&$\bold{1}$&$\bold{1}$&$\bold{1}$&$\bold{1}$ &$\bold{1}$  &$\bold{1}$ & $\bold{1}$ & $\bold{1}$  \\
$U(1)_A$& 4   &$\frac{3}{2}$&-$\frac{3}{2}$&-$\frac{5}{2}$&-3&2&-4&0&7  &9 & -1 & 4  \\
$Z_6$ & 0 & 0 & 1 & 0 & 0 & 0 & 5 & 0 & 3 & 3  & 3 & 3   \\
\bottomrule
\\
\end{tabular}
\begin{tabular}{ccccc}
\toprule
 &$Z_0$&$Z_3$&$Z_4$ &$S$ \\
\midrule
$E_6$ & $\bold{1}$& $\bold{1}$ & $\bold{1}$ & $\bold{1}$ \\
$SU(2)_F$& $\bold{1}$ & $\bold{1}$ & $\bold{1}$ & $\bold{1}$ \\
$U(1)_A$& -1 & -2 & -5 & 9 \\
$Z_6$ & 0 & 3 & 4 & 0 \\
\bottomrule
\end{tabular}
\end{center}
\caption{Field contents and charge assignment under E$_{\text 6}\times$SU(2)$_{\text F} \times$U(1)$_{\text A}\times$Z$_{\text 6}$}
\label{tb:z6complete} 
\end{table}

We introduce the  following fields, which are  listed in Table \ref{tb:z6complete}.
First of all, $\Psi^{ {\bold{27}}}$ is a matter field, and we make its first two generations as a doublet ($\Psi_a^{ {\bold{27}}}$) and the third generation as a singlet ($\Psi_3^{{\bold{27}}}$) of the SU(2)$_{\text F}$, respectively.
The SU(2)$_{\text F}$ and CP are simultaneously broken by the VEVs of flavon fields $F_a$ and $\bar F^a$ as in Eq.\eqref{<F_a>}.
%
%
All the other fields are singlets under SU(2)$_{\text F}$.
$H^{{\bold{27}}}$ is the field whose VEV breaks E$_{\text 6}$ into SO(10), and $\bar H^{\overline {\bold{27}}}$ is introduced to maintain the D-flatness condition.
$C^{ {\bold{27}}}$ is the field whose VEV breaks SO(10) into SU(5), and $\bar C^{\overline {\bold{27}}}$ is introduced to maintain the corresponding D-flatness condition.
%
%
$A^\bold{78} $ and ${A'}^\bold{78} $ are the adjoint fields.
Here, the F-flatness conditions with respect to ${A'}^\bold{78} $ make $A^\bold{78} $ acquire DW-type VEVs for the SO(10) adjoint component to solve the DTS problem \cite{Dimopoulos:1981xm}.
$C'^{ {\bold{27}}}$, $\bar C'^{\overline {\bold{27}}}$, and $Z$ are introduced to give masses for the NG modes and to maintain DW-type VEVs \cite{Barr:1997hq}.

In Table \ref{tb:z6complete}, U(1)$_\text{A}$ charges are assigned so that the DTS and appropriate Yukawa hierarchies are ensured.
Also, $Z_6$ charges are determined so that the requirement discussed in Sec.\ref{subsec:susy cp problem and discrete symmetry} can be fulfilled.
A more detailed discussion for the symmetry breaking in this  model can be found in \cite{Maekawa:2001uk}.
As a result of the SUSY-zero mechanism and $Z_6$ discrete symmetry, possible forms of Yukawa couplings are characteristically restricted.
In the following subsections, we examine consequent Yukawa couplings and derive some predictions of the model.

\subsection{Yukawa couplings}
\label{subsec:yukawa couplings}

In this subsection, we examine Yukawa couplings of the model restoring $ \mathcal O(1)$ coefficients.

Under the charge assignment of Table \ref{tb:z6complete}, the following interactions between matter and Higgs fields are allowed.
\begin{eqnarray}
Y_H &:&
\left( 
\begin{array}{ccc}
0   & d \Psi^a (A,Z_3) \Psi_a & 0 \\
d \Psi^a (A,Z_3) \Psi_a & c \lambda^{2 (\psi_a + \bar f)} \bar F^a \Psi_a \bar F^b \Psi_b & b \lambda^{\psi_a + \psi_3 + \bar f} \bar F^a \Psi_a  \Psi_3 \\
0   & b \lambda^{\psi_a + \psi_3 + \bar f} \Psi_3 \bar F^a \Psi_a & a \lambda^{2 \psi_3 } \Psi_3  \Psi_3
\end{array}
\right) \lambda^h H
\label{eq:yhz6}
\end{eqnarray}
\begin{eqnarray}
Y_C &:&
\left( 
\begin{array}{ccc}
0  & f \lambda^{2 \psi_a + f+ \bar f} F^a \Psi_a \bar F^b \Psi_b & g \lambda^{\psi_a + \psi_3 + f} F^a \Psi_a \Psi_3 \\
f \lambda^{2 \psi_a + f+ \bar f} \bar F^a \Psi_a F^b \Psi_b & 0 & 0 \\
g \lambda^{\psi_a + \psi_3 + f } \Psi_3 F^a \Psi_a & 0 & 0
\end{array}
\right) \lambda^c C \nonumber \\
\label{eq:ycz6}
\end{eqnarray}
These interactions are responsible for the generation of $Y_H$ and $Y_C$.
Here we explicitly write in $\cal O$(1) coefficients, $a,b,c,d,g$ and $f$, for later discussion.
In Eq.\eqref{eq:yhz6}$, \Psi^a (A,Z_3) \Psi_a H$ includes $\lambda^{2 \psi_a + a+z_3+h } \Psi^a Z_3 A \Psi_aH$, $\lambda^{2 (\psi_a +a) +h } \Psi^a A^2 \Psi_a H $, etc. 
Since $A$ acquires $B$-$L$ conserving VEVs ($\langle A \rangle \sim Q_{B-L} \lambda$), the effects of $d \Psi^a \langle (A,Z_3) \rangle \Psi_a H$ differ for the different components of $\Psi_a^\bold{27}$, as we reparametrize below.
Note that all the $\cal O$(1) coefficients are assumed to be real numbers because of the original CP symmetry.

When the Higgs fields and flavon fields acquire VEVs, Eqs.\eqref{eq:yhz6} and \eqref{eq:ycz6} induce the following mass matrix for ${\bold 5}_{i}$, $\bar {\bold 5}_{i}'$ and $\bar {\bold 5}_{i}$:
\begin{eqnarray}
\bordermatrix{ 
   & \bar {\bold{ 5 }}^\prime_1 &\bar {\bold{ 5 }}^\prime_2 & \bar {\bold{ 5 }}^\prime_3 & {\bar {\bold{ 5 }}}_1 &{\bar {\bold{ 5 }}}_2 &{\bar {\bold{ 5 }}}_3\cr
\bold{ 5 }_1& 0  & \alpha d_5 \lambda^5 & 0           & 0             & f e^{i\delta} \lambda^{5.5} & g e^{i\delta} \lambda^{3.5} \cr
\bold{ 5 }_2& -\alpha d_5 \lambda^5 & c \lambda^4 & b \lambda^2 & f e^{i\delta} \lambda^{5.5} & 0  & 0 \cr
\bold{ 5 }_3& 0 & b \lambda^2 & a & g e^{i\delta} \lambda^{3.5} & 0 & 0 \cr}
\langle H \rangle.
\label{eq:z655bp5b}
\end{eqnarray}
Here each power of $\lambda$ is determined by the corresponding U(1)$_{\text A}$ charge.
It is important to note that $\alpha=1$ for the colored Higgs components ($H^C, \bar H^{\bar C}$)  of  $\bold{ 5 }$ and $ \bar {\bold{ 5 }}^\prime_{}$, and  $\alpha=0$ for the doublet Higgs  components ($H_u, H_d$) of $\bold{ 5 }$ and $ \bar {\bold{ 5 }}^\prime_{}$, since the (1,2) and (2,1) elements of Eq.\eqref{eq:z655bp5b}  originate from the B-L conserving VEV of $A$.%
\footnote
{This is true even if higher dimensional term, like $Z_3 \Psi^a A^3 \Psi_a H $, is taken into account.
In terms of SU(4)$\times$SU(2)$_{\text L}\times$SU(2)$_{\text R}$, $H$ and $A$ acquire VEVs in components of $H(\bold{1},\bold{1},\bold{1})$ and  $A(\bold{15},\bold{1},\bold{1})$, respectively.
Therefore two $\Psi_a(\bold{1},\bold{2},\bold{2})$ need to make singlet by themselves in $Z_3 \Psi^a A^3 \Psi_a H $, but anti-symmetric contractions respect to SU(2)$_{\text F}$, SU(2)$_{\text L}$ and SU(2)$_{\text R}$ indices end up zero.
}
Then, we take the following conventions
$d \Psi \langle (Z, A) \rangle \Psi H \supset $
$d_5 \lambda^5 \epsilon ^{a b} H^C[\Psi_a] \bar H^{\bar C}[\Psi_b] {\bold{1}'}[H]$, $\frac{-1}{2} d_q \lambda^5 \epsilon^{ab} Q[\Psi_a]  q_R^C[\Psi_b] H_u[H]$(or $H_d[H]$), $\frac{-3}{2} d_l  \lambda^5 \epsilon^{ab} e_R^C[\Psi_a] L[\Psi_b] H_d[H]$, 
where $\epsilon^{12}=-\epsilon^{21}=1$, $\epsilon^{11}=\epsilon^{22}=0$ and $d_5, d_q, d_l$ are the $ \mathcal O(1)$ coefficients that are different from each other in general.
In Eq.\eqref{eq:z655bp5b}, since $\alpha =0$ for $H_u$ and $ H_d$ components of $\bold{ 5 }$ and $ \bar {\bold{ 5 }}^\prime_{}$, the $H_d[\bar {\bold 5}'_1$] component becomes a purely massless mode.
In other words, $L[\bar {\bold 5}_3$] does not contain a massless mode whose main mode is $H_d[\bar {\bold 5}'_1$].
This fact requires the down-type Higgs doublet to be composed of not only $H_d[H^{ \bar {\bold{5}}^\prime }$] but also $ L $[$C^{{ \bar {\bold{ 5}}}}$]; otherwise the determinant of $Y_E$ vanishes, as we will see later.
%
%

Now we derive the Yukawa couplings $Y_U$, $Y_D$, and $Y_E$ in a semianalytical way, assuming that each $\cal O$(1) coefficient does not largely alter the hierarchical structures that originate in Eqs.\eqref{eq:yhz6}, \eqref{eq:ycz6} and \eqref{eq:z655bp5b}.
First of all, $Y_U$ is simply derived from Eq.\eqref{eq:yhz6} by extracting the corresponding components as\footnote{See \ref{footnote:Higgs mixing} and Appendix \ref{sec:higgs sector}.}%
\begin{eqnarray}
Y_U =
\bordermatrix
{ 
               & {U_{R}^C}_1      & {U_{R}^C}_2      & {U_{R}^C}_3 \cr
Q_{1} & 0 & -\frac{1}{2} d_q \lambda^{5} & 0 \cr
Q_{2} & \frac{1}{2} d_q \lambda^5 & c \lambda^{4} & b \lambda^{2} \cr
Q_{3} & 0 & b \lambda^{2} & a \cr
}.
\label{eq:z6Yu}
\end{eqnarray}
For $Y_D$ and $Y_E$, we first derive the relation between the gauge eigen modes and the mass eigen modes of ${\bar{\bold{5}}}$ from Eq.\eqref{eq:z655bp5b}.
Then we replace $\bar {\bold 5}_{i}$ and $\bar {\bold 5}_{i}'$ of $\Psi_a$ and  $\Psi_3$, whic appeared in Eqs.\eqref{eq:yhz6} and \eqref{eq:ycz6}, in terms of massless modes.
In the same step, we also replace $H_d[\bar {\bold 5}_{H}'$] and $L[\bar {\bold 5}_{C}$] by $H_D$ and $\beta_H \lambda^{0.5} H_D$, respectively.
Here $\beta_H$ denotes another $\cal O$(1) coefficient which enters in this step.
When we leave only the leading order contributions, this procedure leads to the following Yukawa couplings:
\begin{align}
&Y_D = \nonumber \\
&\bordermatrix
{ 
               & {D_{R}^C}_1      & {D_{R}^C}_2      & {D_{R}^C}_3 \cr
Q_{1} & -\{(\frac{bg-af}{g})^2 \frac{1}{ac-b^2} +1 \}\frac{g}{a} g \beta_H e^{2 i\delta} \lambda^{6} & -\frac{bg-af}{g} \frac{d_5}{ac-b^2} g \beta_H e^{i\delta} \lambda^{5.5}& -\frac{1}{2} d_q \lambda^5 \cr
Q_{2} & (\frac{d_q}{2}- \frac{d_5}{ac-b^2} \frac{bg-af}{g} b) \lambda^5 & (-\frac{ad_5^2}{ac-b^2} \frac{b}{g} e^{-i\delta} + \beta_H f e^{i\delta}) \lambda^{4.5} & (\frac{ac-b^2}{a} + \frac{bg-af}{g} \frac{b}{a}) \lambda^{4} \cr
Q_{3} & -\frac{ad_5}{ac-b^2} \frac{bg-af}{g} \lambda^3 & (-\frac{ad_5^2}{ac-b^2} \frac{a}{g} e^{-i\delta} + \beta_H g e^{i\delta}) \lambda^{2.5} & \frac{bg-af}{g} \lambda^{2} \cr
},
\nonumber \\
\label{eq:z6Yd}
\end{align}
\begin{eqnarray}
Y_E^T =
\bordermatrix
{ 
               & L_1      & L_2      & L_3 \cr
{E_R^C}_{1} & -\{(\frac{bg-af}{g})^2 \frac{1}{ac-b^2} +1 \}\frac{g}{a} g \beta_H e^{2i\delta} \lambda^{6} & 0 & -\frac{3}{2} d_l \lambda^5 \cr
{E_R^C}_{2} & \frac{3}{2} d_l \lambda^5 & \beta_H f e^{i\delta} \lambda^{4.5} & (\frac{ac-b^2}{a} + \frac{bg-af}{g} \frac{b}{a}) \lambda^{4} \cr
{E_R^C}_{3} & 0 & \beta_H g e^{i\delta} \lambda^{2.5} & \frac{bg-af}{g} \lambda^{2} \cr
}.
\nonumber \\
\label{eq:z6Ye}
\end{eqnarray}

As a result of discrete symmetry, Eq.\eqref{eq:z6Yu} leads to $Y_u \sim (\frac{d_q}{2})^2 \lambda^6$, where $d_q $ is proportional to a $ Q_B$ charge.%
\footnote{
We assume that this nature is maintained even if higher dimensional terms, like $Z_3 \Psi^a A^3 \Psi_a H $, is taken into account.
In terms of SU(4)$\times$SU(2)$_{\text L}\times$SU(2)$_{\text R}$, SU(4) indices of $\Psi_a(\bold{4},\bold{2},\bold{2})$ can not be contracted with that of $\Psi_a(\bar {\bold{4}},\bold{2},\bold{2})$, because of the SU(2)$_{\text F}$ symmetry.
Then, it is expected that quark components among $\Psi_a(\bold{4},\bold{2},\bold{2})$ and $\Psi_a(\bar {\bold{4}},\bold{2},\bold{2})$ pick up at least two $Q_B$ charges from the contraction with the adjoint fields $A(\bold{15},\bold{1},\bold{1})$.
Strictly speaking, VEVs of $\bold{15}$-plets that are made by the direct product of adjoint fields can give contributions which are not proportional to $Q_B$ for quark components of $(\Psi_a(\bold{4},\bold{2},\bold{2}) \times \Psi_a(\bar {\bold{4}},\bold{2},\bold{2}))_\bold{15}$.
However, these contributions generally possess another suppression factors that are determined so that irreducible representations do not contain a singlet component.}
Also, $Y_U$ turns out to be a real parameters, contrary to $Y_D$ which possesses nonremovable phases that are converted to the KM phase.
In Eq.(\ref{eq:z6Ye}), one can see that the determinant vanishes when $\beta_H$ is set to zero, i.e., $H_D \sim H_d[H^{ \bar {\bold{5}}^\prime }$].
Also note that nonremovable phase in Eq.\eqref{eq:z6Yd} vanishes when $\beta_H$ is set to zero.
This is the reason why we assigned U(1)$_\text{A}$ and $Z_6$ charges so that the down-type Higgs doublet is composed as $H_D \sim H_d[H^{ \bar {\bold{5}}^\prime }$]+$\lambda^{0.5} L $[$C^{{ \bar {\bold{ 5}}}}$].

\subsection{Model analysis}
\label{subsec:model analysis}

The resultant Yukawa couplings of Eqs.\eqref{eq:z6Yu}, \eqref{eq:z6Yd}, \eqref{eq:z6Ye} have rather restricted forms and small numbers of $ \mathcal O(1)$ coefficients.
In this subsection, we examine these Yukawa couplings and derive some predictions.
Also, we discuss the compatibility with observables.

%
First of all, in a semianalytical derivation, where we collect leading contributions in  each element, Eqs.\eqref{eq:z6Yu} and \eqref{eq:z6Yd} lead to the following Cabibbo-Kobayashi-Maskawa (CKM) matrix elements.
\begin{eqnarray}
V_{CKM}^\text{leading} =
\begin{pmatrix} 
1 &  \frac{ a (bg-af)^2  }{2(ac-b^2)  } \frac{ (d_q + 2 d_5) \beta_H }{ \{ (bg-af)^2+(ac-b^2)g^2 \} \beta_H -a^2 d_5^2 e^{2 i \delta}  } \lambda & 0 \\
- V_{12}^* & 1 & \frac{g}{a} \frac{ ac-b^2 }{ bg-af }  \lambda^2 \\
V_{12}^* V_{23}  & - V_{23} & 1
\end{pmatrix}
\label{eq:e6 ckm}
\end{eqnarray}
Here, $V_{ub}$ vanishes at the order of $ \mathcal O(\lambda^3)$ and appers at the next order in $\lambda$ in this model.
It is interesting to note that the current global fit for the CKM matrix elements at low-energy \cite{Amsler:2008zz} or the high-energy extrapolation \cite{Ross:2007az} indeed suggest this type of hierarchical structure.
%
%

Also, in the leading order analysis, we found that  Eqs.\eqref{eq:z6Yu} and \eqref{eq:z6Yd} lead to the following relation,
\begin{eqnarray}
\abs{ V_{cb}  Y_b }= \abs {Y_c}
\label{eq: vcb yb = yc}
\end{eqnarray}
where $Y_b$ and $Y_c$ denote bottom and charm quark Yukawa couplings.
We can guess the origin of this relation as follows.
The (23) element of $V_{CKM}$ is given as $V_{23} \sim \frac{Y_{D23}}{Y_{D33}} - \frac{Y_{U23}}{Y_{U33}}$, but now, $Y_{D23} \sim Y_{U22}$ and $Y_{D33} \sim Y_{U32}$ because ${\bar{\bold{5}}}_2$ turns into ${\bar{\bold{5}}}_3$, after ${\bar{\bold{5}}}$ mixing.
Then $V_{23} Y_b \sim Y_{D23} - \frac{ Y_{U23} }{ Y_{U33} } Y_{D33}  \sim \frac{Y_{U22} Y_{U33} - Y_{U23} Y_{U32}  }{ Y_{U33} }$ and this is roughly $Y_c$.
For comparison with the observables, we quote the results of Ref.\cite{Ross:2007az},
\begin{eqnarray}
| V_{cb} Y_b | \sim \frac{10}{6} | Y_c | 
\label{eq:Ross}
\end{eqnarray} 
where low-energy inputs of $V_{cb}$, $Y_b$ and $Y_c$ are extrapolated to the GUT scale.
Since Eq.\eqref{eq:Ross} is derived in the case of $\tan\beta$=10, Eq.\eqref{eq: vcb yb = yc} fixes $\tan\beta\sim 6$ in the current model%
\footnote{
Since \cite{Ross:2007az} uses the MSSM renormalization group equations (RGE), the discussion can be slightly changed once heavy fields contributions, that are relevant near the GUT scale, are taken into account.
}

%
%
%
%
%

Finally, as an example, we numerically calculate the eigenvalues of quark and lepton Yukawa couplings, a mixing matrix, and a Jarlskog invariant ($\text{J}_\text{CP}$), substituting a set of trial values for the $ \mathcal O(1)$ coefficients  that appeared in Sec.5.2.
We set $a=0.6,\, b=-0.5,\, c=-.7,\, d_5 =-0.9, \,d_q=0.4,\, d_l=-0.5,\,  f=1.5,\, g=-0.9,\, \beta_H= 0.9,\, \delta=1.4$, and repeat the processes described in Sec.5.2.
We obtain the following eigenvalues of the quark and lepton Yukawa couplings, a mixing matrix, and  a Jarlskog invariant.
\begin{eqnarray}
\begin{array}{c}
Y_t = 6(5) \times 10^{\text{-1}} \\
Y_c = 3(1) \times 10^{\text{-3}} \\
Y_u = 4(3) \times 10^{\text{-6}}
\end{array}
\quad
\begin{array}{c}
Y_b = 2(3) \times 10^{\text{-2}} \\
Y_s = 5(6) \times 10^{\text{-4}} \\
Y_d = 8(3) \times 10^{\text{-5}}
\end{array}
\quad
\begin{array}{ccc}
Y_\tau = 3(4) \times 10^{\text{-2}} \\
Y_\mu = 1(3) \times 10^{\text{-3}} \\
Y_e = 3(1)   \times 10^{\text{-5}}
\end{array}
\label{eq:z6massspectra}
\end{eqnarray}
\begin{eqnarray}
|V_{CKM}| = \left(
\begin{array}{ccc}
1            & 2(2) \times 10^{-1} & 2(4) \times 10^{\text{-3}} \\
2(2) \times 10^{-1} & 1    & 10(4) \times 10^{\text{-2}} \\
20(7) \times 10^{-3} & 10(4) \times 10^{-2} & 1
\end{array}
\right)
\label{eq:z6ckm}
\end{eqnarray}
\begin{eqnarray}
\text{J}_\text{CP}= 1(3) \times 10^{\text{-5}}
\label{eq:z6jinv}
\end{eqnarray}
Here the parenthetic digits are the corresponding values of Ref.\cite{Ross:2007az} that are extrapolated to the GUT scale from the low-energy inputs using the MSSM two-loop RGEs%
\footnote{
For the eigenvalues of down-quark and charged lepton Yukawa couplings, that are originally calculated in $\tan \beta =$10, we naively multiply $6/10$ to assess the corresponding values in case of $\tan \beta$=6.
}.
From Eqs.\eqref{eq:z6massspectra}-\eqref{eq:z6jinv}, we see that the current model has the capability to reproduce the quark and lepton mass spectra, mixing angles, and a Jarlskog invariant.

\section{Summary and discussion}
\label{sec:summary and discussion}

In this paper we discussed a model of spontaneous CP violation in the E$_{\text 6}$ SUSY GUT with SU(2)$_{\text F}$ flavor and anomalous U(1)$_{\text A}$ symmetries.

%
We made a model where CP symmetry is spontaneously broken in the flavor sector in order to provide the origin of the KM phase and to evade the SUSY CP problem.
However, as we saw in Sec.\ref{subsec:spontaneous CP violation and mu-term generation}, a naive construction of the model generally leads to an unwanted outcome, Arg$[\mu b^*]=\cal O$(1), when the CP violating effect is taken into account.
Then in Sec.\ref{subsec:susy cp problem and discrete symmetry}, we introduced a discrete symmetry in order to cure this difficulty.
Interestingly, this discrete symmetry plays additional roles.
It ensures that realistic up-quark mass and Cabibbo angle are simultaneously realized without cancellation between $ \mathcal O(1)$ coefficients.
Also, severe constraints from the chromo-electric dipole moment of quark can be satisfied without destabilizing the weak scale.

The discrete symmetry reduces the number of free  parameters, but the model is capable of reproducing quark and lepton mass spectra, mixing angles,  and a Jarlskog invariant, as we see in Eqs.\eqref{eq:z6massspectra}-\eqref{eq:z6jinv}.
We also obtain the characteristic predictions $V_{ub} \sim \cal O$($\lambda^4$) and $| V_{cb} Y_b | = |Y_c|$ at the GUT scale.
Note that, for the trial set of $\cal O$(1) coefficients that is used to calculate Eqs.\eqref{eq:z6massspectra}-\eqref{eq:z6jinv}, Arg[Det[$Y_U Y_D $]] is $\mathcal O(1)$ and it leads to the strong CP phase.
However, this phase can be set to zero by the VEV of the axion that originates from the U(1)$_{\text A}$ breaking.

A more precise fitting of the $ \mathcal O(1)$ coefficients is beyond the scope of the current study, since it requires detailed study of RGEs and threshold corrections.
Note that, in the model, there are many superheavy fields which affect the RGEs.
In particular, among the Higgs fields listed in Table \ref{tb:z6complete}, positively charged fields $C'$, $\bar C'$, and $A'$ provide relatively light fields and these fields start to influence the flow of the gauge couplings from the intermediate scales, as depicted in Fig.1 of the third paper of Ref.\cite{Bando:2001bj}.
Then, this also results in a change of the flows of the Yukawa couplings.

In Sec.\ref{sec:a modle of scpv in e6 susy gut}, we discussed the Yukawa couplings of quarks and charged leptons.
As for the neutrino sector, the left-handed neutrino masses are induced via the see-saw mechanism \cite{seesaw}.
Here, the right-handed neutrino mass terms are obtained by the following interactions,  $\Psi_{i}^{\bold {27}} \Psi_{j}^{\bold {27}} \bar H^{\overline {\bold {27}}} \bar H^{\overline {\bold {27}}}$, $\Psi_{i}^{{\bold {27}}} \Psi_{j}^{ {\bold {27}}} \bar H^{\overline {\bold {27}}} \bar C^{\overline {\bold {27}}}$, $\Psi_{i}^{ {\bold {27}}} \Psi_{j}^{ {\bold {27}}} \bar C^{\overline {\bold {27}}} \bar C^{\overline {\bold {27}}}$, adding appropriate fields like $F_a, \bar F^a, C^{ {\bold {27}}}$ and $\bar C^{\overline {\bold {27}}}$ to make these terms singlets under the symmetries listed in Table \ref{tb:z6complete}.
Then, when we follow the procedure of Ref.\cite{Bando:2001bj}, the solar and atmospheric neutrino mass squared differences are roughly derived as  $\Delta m^2_{\odot }=\cal O$($10^{\text{-6}}) \text{eV}^2$ and $\Delta m^2_{\text{atm}}=\cal O$($10^{\text{-3}}) \text{eV}^2$, respectively.
Note that actual values can change by an order of magnitude, since several new $ \mathcal O(1)$ coefficients appear in the neutrino sector.
Therefore, $\Delta m^2_{\odot }$ and $\Delta m^2_{\text{atm}}$ can be consistent with the current experimental constraints \cite{Amsler:2008zz}.
This is the same for the neutrino mixing angles.
In the current model, the Maki-Nakagawa-Sakata (MNS) matrix  elements \cite{Maki:1962mu} are roughly given as
\begin{eqnarray}
|V_{MNS}| \sim \left(
\begin{array}{ccc}
\lambda^{0.5} & 1                     & \lambda \\
1                     & \lambda^{0.5} & \lambda^{0.5} \\
\lambda^{0.5} & \lambda          & 1
\end{array}
\right),
\label{eq:z6ckm}
\end{eqnarray}
and these values may vary by an order of magnitude when we replace $ \mathcal O(1)$ coefficients with real numbers.
Also, for example, when we change the U(1)$_\text{A}$ charges as $\frac{1}{2} \{c-\bar c -(h-\bar h) \}=-0.5$, $|(V_{MNS})_{13}|$ changes to a more realistic value of $ \mathcal O(\lambda^{1.5})$.
Finally, it is interesting to note that, in the model described in Sec.\ref{sec:a modle of scpv in e6 susy gut}, one of the left-handed neutrino becomes massless.

\section*{Acknowledgement}

S.K is supported in part by Grants-in-Aid for JSPS fellows.
N.M is supported in part by Grants-in-Aid for Scientific Research from the MEXT of Japan.
K.S is supported by the U.K. Science and Technology Facilities Council.
This work was partially supported by the Grant-in-Aid for Nagoya University Global COE Program, ``Quest for Fundamental Principles in the Universe: from Particles to the Solar System and the Cosmos," from the MEXT of Japan.

\appendix

\section{Higgs sector}
\label{sec:higgs sector}

In this section, we briefly outline the Higgs sector of the model that is made to leave $H_u[\bold{10}_H]$ and $H_d[\bold{10}_H] + \lambda^{0.5} L[\bold{16}_C]$ massless. These components are identified as the MSSM Higgs doublets $H_U$ and $H_D $, respectively.
As stressed in Sec.5.2, $H_D \sim H_d[\bold{10}_H] + \lambda^{0.5} L[\bold{16}_C]$ is crucial for the nonvanishing KM phase and electron mass.

In order to distinguish the pair of low-energy massless modes that originate from the Higgs sector, here we compose the operator matrix $O$ which induces mass terms among the SU(2)$_\text{L}$ doublet components of $H,\bar H,C,\bar C,C',\bar C',A$ and $A'$%
\footnote{
See \cite{Maekawa:2001uk} for more detailed discussion.
}.
$O$ consists of 16 rows and 16 columns.
In the following, we take a base where indices are in the order $\bold{10}_{H}$, $\bold{10}_{C} $, $\bold{16}_{C}$, ${\bold{16}}_{A}$, ${\bold{10}}_{C'}$, ${\bold{10}}_{\bar C'}$, ${\bold{16}}_{C'}$, ${\bold{16}}_{A'}$, ${\bold{10}}_{\bar H}$, ${\bold{10}}_{\bar C}$, ${\bold{16}}_{H}$ for rows and $\bold{10}_{H}, \bold{10}_{C} $, $\overline{\bold{16}}_{\bar C}$, $\overline{\bold{16}}_{A}$, ${\bold{10}}_{C'}$, ${\bold{10}}_{\bar C'}$, $\overline{\bold{16}}_{\bar C'}$, $\overline{\bold{16}}_{A'}$, ${\bold{10}}_{\bar H}$, ${\bold{10}}_{\bar C}$, $\overline{\bold{16}}_{\bar H}$ for columns, respectively.
Here we divide $O$ into the following 9 blocks:
\begin{eqnarray}
O =
\bordermatrix{ 
 & \text{\small{$\bold{10}_{H}$\dots}} & \dots & \text{\small{\dots $\overline{\bold{16}}_{\bar H}$}}\cr
\text{\quad \rotatebox[origin=c]{-90}{\small{\rotatebox[origin=c]{90}{$\bold{10}_{H}$} $\dots$}}} & \bold{0}_{4 \times 4} & A_{4 \times 4} & \bold{0}_{4 \times 3} \cr
\quad \,\,\, \text{\rotatebox[origin=c]{-90}{\small{\quad $\dots$ \quad}}} & B_{4 \times 4} & C_{4 \times 4} & D_{4 \times 3} \cr
\text{\quad \rotatebox[origin=c]{-90}{\small{$\dots$ \rotatebox[origin=c]{90}{$\bold{16}_{H}$}}}} & E_{3 \times 4} & F_{3 \times 4} & G_{3 \times 3} \cr
}
\label{eq:operator matrix}
\end{eqnarray}
where subscripts indicate the number of rows and columns of the partial matrices.
Here the $\bold{0}$'s have no entries and $E$ has a few entries since the corresponding fields are mostly negatively charged under U(1)$_\text{A}$ (SUSY-zero mechanism).
On the contrary, the elements of $C, D, F$ and $G$ are almost filled since the corresponding fields are mostly positively charged.
Therefore, in the following, we concentrate on $A$ and $ B$ which determine the composition of the massless modes.
%
%
Here $B$ is responsible for the composition of $H_U$ and $A$ is responsible for the composition of $H_D$.

Under the charge assignment given in Table.\ref{tb:z6complete}, $A$ and $B$ are filled with the following terms.
\begin{align}
A = 
\bordermatrix{
 &\bold{10}_{C'}&\bold{10}_{\bar C'}&\overline{\bold{16}}_{\bar C'} & \overline{\bold{16}}_{A'} \cr
\bold{10}_H & H^2 A C' & 0 & HH \bar C'\bar C & 0 \cr
\bold{10}_C & 0 & \bar C'(AF\bar F +Z_4)C & 0 & 0 \cr
\bold{16}_C & 0 & 0 & \bar C'(AF\bar F+Z_4)C & 0 \cr
\bold{16}_A & 0 & \bar C'AC & \bar C'AH &A'A \cr
} 
\label{eq:aigen}
\end{align}
\begin{eqnarray}
B =
\bordermatrix{ 
 &\bold{10}_{H}&\bold{10}_{C}&\overline{\bold{16}}_{\bar C} & \overline{\bold{16}}_{A} \cr
\bold{10}_{C'} & H^2 A C'& 0 &C'H (A+Z_3)\bar C\bar C& \bar CAC'Z_3 \cr
\bold{10}_{\bar C'} &0& \bar C'(AF\bar F+Z_4)C& \bar C'(A+Z_3) \bar C^2 & \bar C'A\bar C \bar H Z_3 \cr
\bold{16}_{C'} & 0 & 0 & 0 & \bar H A C'\cr
\bold{16}_{A'} & 0 & 0 & 0 & A'A \cr
}
\label{eq:bigen}
\end{eqnarray}
Then, after symmetry breaking, Eqs.\eqref{eq:aigen} and \eqref{eq:bigen} induce the following mass terms among SU(2)$_\text{L}$ components:
\begin{eqnarray}
A =
\bordermatrix{ 
%
&\bold{10}_{C'}&\bold{10}_{\bar C'}&\overline{\bold{16}}_{\bar C'} & \overline{\bold{16}}_{A'} \cr
\bold{10}_H & 0 & 0 & \lambda^{5.5} & 0 \cr
\bold{10}_C & 0 & \lambda^{5.5} & 0 & 0 \cr
\bold{16}_C & 0 & 0 & \lambda^{5} & 0 \cr
\bold{16}_A & 0 & \lambda^{6.5} & \lambda^{5.5} & \lambda^{3} \cr
} 
\label{eq:al-hhbar-1r1/2}
\end{eqnarray}
\begin{eqnarray}
B_L =
\bordermatrix{ 
%
&\bold{10}_{H}&\bold{10}_{C}&\overline{\bold{16}}_{\bar C} & \overline{\bold{16}}_{A} \cr
\bold{10}_{C'} & 0 & 0 & \lambda^{8.5} & \lambda^9 \cr
\bold{10}_{\bar C'} &0& \lambda^5& \lambda^{13} & \lambda^{12.5}  \cr
\bold{16}_{C'} & 0 & 0 & 0 & \lambda^{8.5} \cr
\bold{16}_{A'} & 0 & 0 & 0 & \lambda^3 \cr
}
\label{eq:bl-hhbar-1r1/2}
\end{eqnarray}
Note that each (1,1) entry vanishes since $A$ acquires $B-L$ conserving VEV.
Therefore, Eq.\eqref{eq:al-hhbar-1r1/2} and \eqref{eq:bl-hhbar-1r1/2} appropriately realize $H_D \sim H_d[\bold{10}_H] + \lambda^{0.5} L[\bold{16}_C]$ and $H_U \sim H_u[\bold{10}_H]$.

\end{document}